%% file: ms.tex
\documentclass[conference,twocolumn,10pt]{IEEEtran} 
\input{sec-preamble.tex}

\newcounter{mytempeqncnt}

\begin{document}

\input{sec-title.tex}

\input{sec-abstract.tex}

\input{sec-peer-review-title.tex}

\glsresetall

\input{sec-introduction.tex}

\input{sec-system-model-v3.tex}

\input{sec-contribution.tex}

\input{sec-simulation-results.tex}

\input{sec-conclusion.tex}


\input{sec-bibliography.tex}

\end{document}

%% file: sec-preamble.tex
\ifCLASSINFOpdf
  \usepackage[pdftex]{graphicx}
  \usepackage{epstopdf} 
\else
\fi
\usepackage{amsmath}
  \usepackage[caption=false,font=footnotesize]{subfig}
\usepackage{url}

\input{mymath.tex}
\input{glossary.tex}

\newcommand{\tabref}[1]{Table~\ref{#1}}
\newcommand{\figref}[1]{\figurename~\ref{#1}}


%% file: mymath.tex
\newcommand{\comment}[1]{{}}

\usepackage{amssymb}
\usepackage{xspace}

\newcommand{\complex}{\ensuremath{\mathbb{C}}\xspace}
\newcommand{\complexpow}[2]{\ensuremath{\complex^{#1 \times #2}}\xspace}

\renewcommand{\j}{\ensuremath{\textrm{j}}}

\usepackage{bm}

\newcommand{\Ts}{\ensuremath{T_{\textrm{s}}}}
\newcommand{\fs}{\ensuremath{f_{\textrm{s}}}}
\newcommand{\cplength}{\ensuremath{N_{\textrm{CP}}}}

\newcommand{\inv}{\ensuremath{^{-1}}\xspace}

\newcommand{\ctrans}{\ensuremath{^{{*}}}\xspace}

\newcommand{\mat}[1]{\ensuremath{\mathbf{#1}}\xspace} 
\renewcommand{\vec}[1]{\ensuremath{\mathbf{#1}}\xspace} 
\newcommand{\matfd}[1]{\ensuremath{\bm{\mathsf{#1}}}\xspace} 
\newcommand{\vecfd}[1]{\ensuremath{\bm{\mathsf{#1}}}\xspace} 

\newcommand{\pnorm}[2]{\ensuremath{\left\lVert{#2}\right\rVert_{#1}}\xspace}

\newcommand{\entry}[2]{\ensuremath{\left[{#1}\right]_{#2}}\xspace}

\newcommand{\svdfd}[2]{\ensuremath{\matfd{U}_{#1}#2 \ \matfd{\Sigma}_{#1}#2 \ \matfd{V}\ctrans_{#1}#2 }}

\newcommand{\lsingfd}[2]{\ensuremath{\matfd{U}_{#1}#2 }}
\newcommand{\rsingfd}[2]{\ensuremath{\matfd{V}_{#1}#2 }}

\newcommand{\cgauss}[2]{\ensuremath{\mathcal{N}_{\complex} \left( {#1} , {#2} \right) }\xspace} 

\newcommand{\prebbfd}{\ensuremath{\matfd{F}_{\mathsf{BB}}}\xspace} 
\newcommand{\prerffd}{\ensuremath{\matfd{F}_{\mathsf{RF}}}\xspace} 
\newcommand{\prebbfdtilde}{\ensuremath{\hat{\matfd{F}}_{\mathsf{BB}}}\xspace} 

\newcommand{\prefd}{\ensuremath{\matfd{F}}\xspace} 
\newcommand{\combbfd}{\ensuremath{\matfd{W}_{\mathsf{BB}}}\xspace} 
\newcommand{\comfd}{\ensuremath{\matfd{W}}\xspace} 
\newcommand{\comrffd}{\ensuremath{\matfd{W}_{\mathsf{RF}}}\xspace} 

\newcommand{\xfd}{\ensuremath{\matfd{X}}\xspace} 
\newcommand{\xbbfd}{\ensuremath{\matfd{X}_{\mathsf{BB}}}\xspace} 
\newcommand{\xrffd}{\ensuremath{\matfd{X}_{\mathsf{RF}}}\xspace} 
\newcommand{\xbbfdtilde}{\ensuremath{\bar{\matfd{X}}_{\mathsf{BB}}}\xspace} 

\newcommand{\Arffd}{\ensuremath{\matfd{A}_{\mathsf{RF}}}\xspace}

\newcommand{\channel}{\ensuremath{\mat{H}}\xspace} 
\newcommand{\channelfd}{\ensuremath{\matfd{H}}\xspace} 

\newcommand{\Nt}{\ensuremath{N_\textrm{t}}\xspace} 
\newcommand{\Nr}{\ensuremath{N_\textrm{r}}\xspace} 
\newcommand{\Na}{\ensuremath{N_{\textrm{a}}}\xspace} 

\newcommand{\Nrf}{\ensuremath{N_\textrm{RF}}\xspace} 
\newcommand{\Nrftx}{\ensuremath{L_{\textrm{t}}}\xspace} 
\newcommand{\Nrfrx}{\ensuremath{L_{\textrm{r}}}\xspace} 

\newcommand{\Ns}{\ensuremath{N_\textrm{s}}\xspace} 

\newcommand{\Rsfd}{\ensuremath{\matfd{R}_{\mathsf{s}}\xspace}}
\newcommand{\Rnfd}{\ensuremath{\matfd{R}_{\mathsf{n}}\xspace}}

\newcommand{\Pt}{\ensuremath{P_{\textrm{tx}}\xspace}}

\newcommand{\symbvecfd}{\ensuremath{\vecfd{s}}\xspace}
\newcommand{\noisevecfd}{\ensuremath{\vecfd{n}}\xspace}
\newcommand{\rxvecfd}{\ensuremath{\hat{\vecfd{s}}}\xspace}

\newcommand{\numrays}{\ensuremath{N_{\textsf{rays}}}\xspace}
\newcommand{\numclust}{\ensuremath{N_{\textsf{clust}}}\xspace}

\newcommand{\frobenius}[1]{\ensuremath{\pnorm{\textsf{F}}{#1}}}

\newcommand{\aoa}{\ensuremath{\theta}\xspace}
\newcommand{\aod}{\ensuremath{\phi}\xspace}

\newcommand{\rxresponsevector}{\ensuremath{\vec{a}_{\textrm{r}}(\azimuth_{i,j},\elevation_{i,j})}\xspace}
\newcommand{\txresponsevector}{\ensuremath{\vec{a}_{\textrm{t}}(\azimuth_{i,j},\elevation_{i,j})}\xspace}

\renewcommand{\rxresponsevector}{\ensuremath{\vec{a}_{\textrm{r}}}\xspace}
\renewcommand{\txresponsevector}{\ensuremath{\vec{a}_{\textrm{t}}}\xspace}

\newcommand{\snr}{\ensuremath{\mathrm{SNR}}\xspace}

\newcommand{\node}[1]{\unskip\ensuremath{^{^{\left(#1\right)}}}\xspace}

\newcommand{\ctransnode}[1]{\ensuremath{^{{*}^{\left(#1\right)}}}\xspace}



%% file: glossary.tex
\usepackage{xspace}
\usepackage[acronym,nogroupskip,nonumberlist,nopostdot]{glossaries}
\makeglossaries

\usepackage{xcolor}

\newacronym{snr}{SNR}{signal-to-noise ratio}
\newacronym{sinr}{SINR}{signal-to-interference-plus-noise ratio}
\newacronym{inr}{INR}{interference-to-noise ratio}
\newacronym{sir}{SIR}{signal-to-interference ratio}
\newacronym{ian}{IAN}{interference as noise}
\newacronym{ber}{BER}{bit error rate}
\newacronym{pn}{PN}{pseudorandom noise}
\newacronym{bfsk}{BFSK}{binary frequency shift keying}
\newacronym{fh}{FH}{frequency-hopped}
\newacronym{fh-bfsk}{FH-BFSK}{frequency-hopped binary frequency shift keying}
\newacronym{crc}{CRC}{cyclic redundancy check}
\newacronym{isi}{ISI}{intersymbol interference}
\newacronym{dsss}{DSSS}{direct-sequence spread spectrum}
\newacronym{ofdm}{OFDM}{orthogonal frequency-division multiplexing}
\newacronym{ofdma}{OFDMA}{orthogonal frequency-division multiple access}
\newacronym{sdr}{SDR}{software-defined radio}
\newacronym{tx}{TX}{transmitter}
\newacronym{rx}{RX}{receiver}
\newacronym{fdd}{FDD}{frequency-division duplexing}
\newacronym{tdd}{TDD}{time-division duplexing}
\newacronym{fdma}{FDMA}{frequency-division multiple access}
\newacronym{tdma}{TDMA}{time-division multiple access}
\newacronym{sdma}{SDMA}{space-division multiple access}
\newacronym[plural=MPCs]{mpc}{MPC}{multipath component}
\newacronym{mui}{MUI}{multi-user interference}

\newacronym{ls}{LS}{least-squares}
\newacronym{lms}{LMS}{least mean squares}
\newacronym{rls}{RLS}{recursive least-squares}
\newacronym{rzf}{RZF}{regularized zero-forcing}
\newacronym{mmse}{MMSE}{minimum mean square error}
\newacronym{lmmse}{LMMSE}{linear minimum mean square error}
\newacronym{mse}{MSE}{mean square error}
\newacronym{fft}{FFT}{fast Fourier transform}
\newacronym{dft}{DFT}{discrete Fourier transform}
\newacronym{dtft}{DTFT}{discrete-time Fourier transform}
\newacronym{ctft}{CTFT}{continuous-time Fourier transform}
\newacronym{ml}{ML}{machine learning}
\newacronym[plural=NNs]{nn}{NN}{neural network}
\newacronym[plural=RNNs]{rnn}{RNN}{recurrent neural network}
\newacronym[plural=ADCs]{adc}{ADC}{analog-to-digital converter}
\newacronym[plural=DACs]{dac}{DAC}{digital-to-analog converter}
\newacronym[plural=FPGAs]{fpga}{FPGA}{field-programmable gate array}
\newacronym{evm}{EVM}{error vector magnitude}
\newacronym{psd}{PSD}{power spectral density}
\newacronym{enob}{ENOB}{effective number of bits}
\newacronym{zf}{ZF}{zero-forcing}
\newacronym{rv}{r.v.}{random variable}
\newacronym{omp}{OMP}{orthogonal matching pursuit}
\newacronym{svd}{SVD}{singular value decomposition}

\newacronym{agc}{AGC}{automatic gain control}
\newacronym{rf}{RF}{radio frequency}
\newacronym{los}{LOS}{line-of-sight}
\newacronym{nlos}{NLOS}{non-line-of-sight}
\newacronym{ple}{PLE}{path loss exponent}
\newacronym[plural=dB]{db}{dB}{decibel}
\newacronym{pa}{PA}{power amplifier}
\newacronym{lna}{LNA}{low noise amplifier}
\newacronym{cw}{CW}{continuous wave}
\newacronym{papr}{PAPR}{peak-to-average power ratio}
\newacronym{usrp}{USRP}{Universal Software Radio Peripheral}
\newacronym{irr}{IRR}{image rejection ratio}
\newacronym{lo}{LO}{local oscillator}
\newacronym{vm}{VM}{vector modulator}
\newacronym{mmwave}{mmWave}{millimeter-wave}

\newacronym{csma}{CSMA}{carrier-sense multiple access}
\newacronym{csmaca}{CSMA/CA}{carrier-sense multiple access with collision avoidance}
\newacronym{csmacd}{CSMA/CD}{carrier-sense multiple access with collision detection}
\newacronym{mac}{MAC}{medium access control}
\newacronym{phy}{PHY}{physical layer}
\newacronym{4g}{4G}{fourth generation}
\newacronym{lte}{LTE}{Long-Term Evolution}
\newacronym{4glte}{4G LTE}{\gls{4g} \gls{lte}}
\newacronym{5g}{5G}{fifth generation}
\newacronym{nr}{NR}{New Radio}
\newacronym{5gnr}{5G NR}{\gls{5g} \gls{nr}}
\newacronym{ieee}{IEEE}{Institute of Electrical and Electronics Engineers}
\newacronym{wifi}{Wi-Fi}{IEEE 802.11}
\newacronym{lan}{LAN}{local area network}
\newacronym{wlan}{WLAN}{wireless local area network}
\newacronym[plural=BSs]{bs}{BS}{base station}
\newacronym[plural=SBSs]{sbs}{SBS}{small-cell base station}
\newacronym[plural=FD-SBSs]{fdsbs}{FD-SBS}{\gls{fd}-enabled \gls{sbs}}
\newacronym[plural=MBSs]{mbs}{MBS}{macrocell base station}
\newacronym[plural=UEs]{ue}{UE}{user equipment}
\newacronym{ul}{UL}{uplink}
\newacronym{dl}{DL}{downlink}
\newacronym{qos}{QoS}{Quality of Service}
\newacronym{fcc}{FCC}{Federal Communications Commission}
\newacronym{iab}{IAB}{integrated access and backhaul}
\newacronym{fab}{FAB}{fixed access and backhaul}
\newacronym{hetnet}{HetNet}{heterogeneous network}

\newacronym{siso}{SISO}{single-input single-output}
\newacronym{mimo}{MIMO}{multiple-input multiple-output}
\newacronym{sumimo}{SU-MIMO}{single-user \gls{mimo}}
\newacronym{mumimo}{MU-MIMO}{multi-user \gls{mimo}}
\newacronym{bf}{BF}{beamforming}
\newacronym{ca}{CA}{constant amplitude}
\newacronym{ula}{ULA}{uniform linear array}
\newacronym{aoa}{AoA}{angle of arrival}
\newacronym{aod}{AoD}{angle of departure}
\newacronym{dof}{DoF}{degrees of freedom}
\newacronym{csi}{CSI}{channel state information}
\newacronym{csit}{CSIT}{\gls{csi} at the transmitter}
\newacronym{csir}{CSIR}{\gls{csi} at the receiver}

\newacronym{fd}{FD}{in-band full-duplex}
\newacronym{hd}{HD}{half-duplex}
\newacronym{si}{SI}{self-interference}
\newacronym{sic}{SIC}{self-interference cancellation}
\newacronym{soi}{SoI}{signal of interest}
\newacronym{asic}{A-SIC}{analog \gls{sic}}
\newacronym{dsic}{D-SIC}{digital \gls{sic}}
\newacronym{star}{STAR}{simultaneous transmit and receive}
\newacronym{warp}{WARP}{Wireless Open-Access Research Platform}
\newacronym{bfc}{BFC}{beamforming cancellation}
\newacronym{ipi}{IPI}{inter-panel-interference}
\newacronym{ipic}{IPIC}{inter-panel-interference cancellation}

\newacronym{qcqp}{QCQP}{quadratically-constrained quadratic programming}
\newacronym{cdf}{CDF}{cumulative density function}

\newacronym{elf}{ELF}{extremely low frequency}
\newacronym{slf}{SLF}{super low frequency}
\newacronym{ulf}{ULF}{ultra low frequency}
\newacronym{vlf}{VLF}{very low frequency}
\newacronym{lf}{LF}{low frequency}
\newacronym{mf}{MF}{medium frequency}
\newacronym{hf}{HF}{high frequency}
\newacronym{vhf}{VHF}{very high frequency}
\newacronym{uhf}{UHF}{ultra high frequency}
\newacronym{shf}{SHF}{super high frequency}
\newacronym{ehf}{EHF}{extremely high frequency}
\newacronym{thf}{THF}{tremendously high frequency}

\newacronym{wncg}{WNCG}{Wireless Networking and Communications Group}
\newacronym{linc}{LINC}{Laboratory of Informatics, Networks, and Communications}
\newacronym{ut}{UT Austin}{The University of Texas at Austin}
\newacronym{uiuc}{UIUC}{University of Illinois at Urbana-Champaign}
\newacronym{usc}{USC}{University of Southern California}
\newacronym{mit}{MIT}{Massachusetts Institute of Technology}
\newacronym{berkeley}{UC Berkeley}{University of California, Berkeley}
\newacronym{osu}{OSU}{Ohio State University}

\newcommand{\mmwave}{\gls{mmwave}\xspace}
\newcommand{\mimo}{\gls{mimo}\xspace}

\newcommand{\rf}{\gls{rf}\xspace}

\newcommand{\fg}{\gls{5g}\xspace}
\newcommand{\fd}{\gls{fd}\xspace}
\newcommand{\hd}{\gls{hd}\xspace}

\newcommand{\si}{\gls{si}\xspace}

\newcommand{\sic}{\gls{sic}\xspace}

\newcommand{\bfc}{\gls{bfc}\xspace}

\newcommand{\zf}{\gls{zf}\xspace}
\newcommand{\lmmse}{\gls{lmmse}\xspace}
\newcommand{\rzf}{\gls{rzf}\xspace}

%% file: sec-title.tex
%
\title{Frequency-Selective Beamforming Cancellation Design for Millimeter-Wave Full-Duplex}
%
%
%

\author{Ian~P.~Roberts, Hardik~B.~Jain, and Sriram Vishwanath\\University of Texas at Austin and GenXComm, Inc.\\ \{ipr, hardikbjain, sriram\}@utexas.edu}

\maketitle

%% file: sec-abstract.tex
\begin{abstract}
The wide bandwidths offered at \mmwave frequencies have made them an attractive choice for future wireless communication systems. Recent works have presented beamforming strategies for enabling \fd capability at \mmwave even under the constraints of hybrid beamforming, extending the exciting possibilities of next-generation wireless. Existing \mmwave \fd designs, however, do not consider frequency-selective \mmwave channels. Wideband communication at \mmwave suggests that frequency-selectivity will likely be of concern since communication channels will be on the order of hundreds of megahertz or more. This has motivated the work of this paper, in which we present a frequency-selective beamforming design to enable practical wideband \mmwave \fd applications. In our designs, we account for the challenges associated with hybrid analog/digital beamforming such as phase shifter resolution, a desirably low number of \rf chains, and the frequency-flat nature of analog beamformers. We use simulation to validate our work, which indicates that spectral efficiency gains can be achieved with our design by enabling simultaneous transmission and reception in-band. 
\end{abstract}

%% file: sec-peer-review-title.tex
%
\IEEEpeerreviewmaketitle

%% file: sec-introduction.tex
\section{Introduction} \label{sec:introduction}

Future wireless networks have turned to \mmwave frequencies due to their wide bandwidths, which can offer higher rates and support more users \cite{Andrews_Buzzi_Choi_Hanly_Lozano_Soong_Zhang_2014}. While there are challenges associated with communication at such high frequencies, researchers have presented creative solutions that enable practical \mmwave communication. Chief among these is hybrid beamforming, where the combination of a digital beamformer and an analog beamformer is used to achieve performance comparable to an unconstrained fully-digital beamformer with a reduced number of \rf chains \cite{Heath_Gonzalez-Prelcic_Rangan_Roh_Sayeed_2016}. The analog beamformer's phase shifter resolution and lack of amplitude control are constraints that often complicate the hybrid approximation process \cite{Heath_Gonzalez-Prelcic_Rangan_Roh_Sayeed_2016}. 

The expected impacts of \mmwave communication could be extended even further by equipping \mmwave systems with \fd capability---the ability for a device to simultaneously transmit and receive in-band. Such a capability offers exciting benefits over traditional \hd operation, where the incurred \si is prohibitive. Most immediately, \fd operation has the potential of doubling the achievable spectral efficiency since orthogonal duplexing of transmission and reception is no longer necessary. Furthermore, at a network level, \fd capability could offer significant throughput gains thanks to the transformative ability to sense the channel while transmitting, suggesting that overhead and challenges associated with conventional medium access techniques can be avoided. Additionally, the relaying associated with \fg network densification is a particularly suitable application for \mmwave \fd where a relay node could simultaneously forward and receive data.

In recent years, \fd has become a reality thanks to the development of various \sic techniques that take advantage of the fact that a transceiver knows its own transmission, allowing it to reconstruct the \si and subtract it at the receiver. If done properly, \sic leaves a desired receive signal virtually free from \si. Almost all existing \fd techniques, however, have been for systems operating at sub-$6$ GHz frequencies and cannot be applied directly to \mmwave systems largely due to the numerous antennas in use and high nonlinearity. For this reason, the few existing \fd designs for \mmwave propose methods for mitigating \si by a means referred to as \bfc where the transmit and receive beamformers at a transceiver are designed to avoid introducing \si \cite{ipr-bfc-2019-globecom}. 

Given that wide bandwidths will be used by \mmwave systems, frequency-selective channels will likely be of concern. 
Existing \bfc designs for \mmwave \fd, however, do not consider this frequency-selectivity, presenting designs for frequency-flat channels \cite{ipr-bfc-2019-globecom,Satyanarayana_El-Hajjar_Kuo_Mourad_Hanzo_2019}. 
While schemes such as \gls{ofdm} can transform a frequency-selective channel into a bank of frequency-flat subchannels, the existing \bfc designs cannot be applied on a subcarrier basis due to the frequency-flat nature of the analog beamformer. 
For this reason, we present a frequency-selective \bfc design to enable wideband \mmwave \fd applications.
Our design captures the hybrid beamforming structure used at \mmwave and incorporates its limitations during the design. We account for phase shifter resolution and a lack of amplitude control commonly associated with \rf beamformers. Most importantly, our design is a frequency-selective one, enabling practical wideband \fd operation at \mmwave.



\input{par-notation.tex}

%% file: par-notation.tex
We use the following notation. 
We use bold uppercase, \mat{A}, to represent matrices. 
We use bold lowercase, \vec{a}, to represent column vectors. 
We use $(\cdot)$\ctrans and \frobenius{\cdot} to represent the conjugate transpose and Frobenius norm, respectively. 
We use $\left[\mat{A}\right]_{i,j}$ to denote the element in the $i$th row and $j$th column of \mat{A}. 
We use $\left[\mat{A}\right]_{i,:}$ and $\left[\mat{A}\right]_{:,j}$ to denote the $i$th row and $j$th column of \mat{A}. 
We use $\cgauss{\vec{m}}{\mat{R}}$ as a multivariate circularly symmetric complex Normal distribution having mean $\vec{m}$ and covariance matrix $\mat{R}$.
We use the term ``beamformer'' most often to refer to a single column (rather than matrix) of beamforming weights.

%% file: sec-system-model-v3.tex
\section{System Model} \label{sec:system-model}


In our design, we consider a \mmwave network of three nodes: $i$, $j$, and $k$. Node $i$ is a \fd transceiver transmitting to $j$ while receiving from $k$ (in-band). Node $i$ is capable of simultaneously realizing independent transmit and receive beamformers at separate arrays. Nodes $j$ and $k$ are \hd devices. We remark that, together, $j$ and $k$ could represent another \fd device. We assume that all precoding and combining is done via fully-connected hybrid beamforming \cite{Heath_Gonzalez-Prelcic_Rangan_Roh_Sayeed_2016}, where the combination of baseband and \rf beamforming is used at each transmitter and receiver.

Let $\Nt\node{m}$ $(\Nr\node{m})$ be the number of transmit (receive) antennas at node $m \in \{i,j,k\}$. Let $\Nrftx\node{m}$ $(\Nrfrx\node{m})$ be the number of \rf chains at the transmitter (receiver) of node $m \in \{i,j,k\}$. 
Let all channels be frequency-selective over the band of interest. Being frequency-selective, assume the channel impulse response has length $D$ taps, and $d \in \{ 0,\dots,D-1 \}$ represents the channel tap index (in time).

For the desired channels $\left\{\channel_{ij}[d]\right\}$ and $\left\{\channel_{ki}[d]\right\}$, we use the model in \eqref{eq:desired-channel}, where the \mimo channel matrix between a node's transmit array and a different node's receive array at tap $d$ as 
\begin{align}
\channel[d] = \alpha \sum_{m=1}^{\numclust} \sum_{n=1}^{\numrays} \beta_{mn} p\left(d \Ts - \tau_{mn} \right) \rxresponsevector(\aoa_{mn}) \txresponsevector\ctrans(\aod_{mn}), \label{eq:desired-channel}
\end{align}
where $\alpha = \sqrt{{\Nt \Nr}/{(\numrays \numclust)}}$, $\rxresponsevector(\aoa_{mn})$ and $\txresponsevector(\aod_{mn})$ are the receive and transmit array responses, respectively, for ray $n$ within cluster $m$, which has some \gls{aoa}, $\aoa_{mn}$, and \gls{aod}, $\aod_{mn}$ \cite{Heath_Gonzalez-Prelcic_Rangan_Roh_Sayeed_2016,frequency-selective-hybrid-heath}. The total number of clusters is $\numclust$ and the number of rays per cluster is $\numrays$.
Each ray has a random gain $\beta_{mn} \sim \cgauss{0}{1}$. The function $p(t)$ captures the gain at time $t$ due to pulse shaping, and $\tau_{mn}$ is the associated relative delay of ray $n$ within cluster $m$. The sampling rate of the channel is $1/\Ts$ samples per second.

For the \si channel at $i$, $\{\channel_{ii}[d]\}$, we use the Rician summation shown in \eqref{eq:si-channel-rice}, where $\kappa$ is the Rician factor. The \gls{los} component $\channel^{\textrm{LOS}}_{ii}$ is based on the spherical-wave \mimo model in \cite{ipr-bfc-2019-globecom,spherical-wave-mimo} and is frequency-flat, reserving it due to space constraints. The \gls{nlos} component $\channel^{\textrm{NLOS}}_{ii}[d]$ is based on the model in \eqref{eq:desired-channel}, meaning it is frequency-selective.
\input{eq-si-channel-rice.tex}
It is worthwhile to remark that the \si channel is not well-characterized for \mmwave \fd systems, meaning our model may not align well with practice. The channel in \eqref{eq:si-channel-rice} is full-rank, though we do not rely on its particular channel structure in our design. We expect our design will translate well to other, more practically-sound frequency-selective \si channels.

We assume \gls{ofdm} is used for sufficiently transforming a frequency-selective wideband channel into a bank of $U$ frequency-flat subchannels by use of cyclic prefix of length $\cplength$. Let $u \in \{0,\dots,U-1\}$ denote the subcarrier index. 
Let $\channelfd_{mn}[u]$ be the frequency-flat \mimo channel matrix from node $m$ to node $n$ on subcarrier $u$ where $m,n \in \{i,j,k\}$.
Having an effectively circularly-convolved channel in time, we take a $U$-point \gls{dft} of the channel matrices across time to get \mimo channels across subcarriers, which are frequency-flat.
\begin{align}
\matfd{H}[u] = \sum_{d=0}^{D-1} \channel[d] \ e^{-\j \frac{2 \pi u d}{U}}
\end{align}




Let $\symbvecfd\node{m}[u] \sim \cgauss{\mat{0}}{\Rsfd}$ be the $\Ns\node{m} \times 1$ symbol vector transmitted on subcarrier $u$ intended for node $m \in \{i,j,k\}$, where we have assumed each subcarrier is transmitted with the same number of streams. 
Let $\noisevecfd\node{m}[u] \sim \cgauss{\mat{0}}{\sigma^2 \Rnfd}$ be a noise vector received by the receive array at node $m$ on subcarrier $u$, where we have assumed a common (and frequency-flat) noise covariance matrix. Defining $\sigma^2$ as the noise variance and letting $\Rnfd = \mat{I}$ and $\Rsfd = \frac{1}{U}\mat{I}$ provides a convenient formulation. 
For $m,n \in \{i,j,k\}$, we define the per subcarrier link \gls{snr} from node $m$ to node $n$ as 
$ \snr_{mn} \triangleq {\Pt\node{m} G^2_{mn}}/({\sigma^2 B})$, 
where $\Pt\node{m}$ is the transmit power at $m$, $G^2_{mn}$ is the large-scale power gain of the propagation from $m$ to $n$, and $B$ is the total bandwidth. Our simulated results in Section~\ref{sec:simulation-results} are not concerned with a particular $B$; we simply use the \gls{snr} as a varied scalar quantity for evaluating our design. 

Let $\prebbfd\node{m}[u]$ $(\combbfd\node{m}[u])$ be the baseband precoding (combining) matrix at node $m \in \{i,j,k\}$ for subcarrier $u$. Unlike the baseband beamformers, the \rf beamformers are not frequency-selective but rather frequency-flat and are not indexed per subcarrier. This fact will impact our design in Section~\ref{sec:contribution}. Let $\prerffd\node{m}$ $(\comrffd\node{m})$ be the \rf precoding (combining) matrix at $m \in \{i,j,k\}$. We impose a constant amplitude constraint on the entries of the \rf beamformers representing the fact that they have phase control and no amplitude control. Further, we assume the phase shifters have finite resolution, though our design does not rely on a particular resolution.

For the sake of analysis, we impose a uniform power allocation across subcarriers and streams. To do this, we normalize the baseband precoder for each stream on subcarrier $u$ such that 
${\Big\Vert\prerffd\node{m}\entry{\prebbfd\node{m}[u]}{:,\ell}\Big\Vert_{\textsf{F}}}^2 = 1 \ \forall \ \ell \in [0,\Ns\node{m}-1]$  
, which ensures that for all $m \in \{i,j,k\}$
\begin{align}
\sum_{u=0}^{U-1} \frobenius{\prerffd\node{m}{\prebbfd\node{m}[u]}}^2 = U \Ns\node{m}.
\end{align}

\begin{figure*}[!t]
	\normalsize
	\setcounter{mytempeqncnt}{\value{equation}}
	\setcounter{equation}{4}
	\begin{align}
	\rxvecfd\node{i}[u] =& \combbfd\ctransnode{i}[u] \comrffd\ctransnode{i} \left(\sqrt{\snr_{ki}} \  \channelfd_{ki}[u] \prerffd\node{k} \prebbfd\node{k}[u] \symbvecfd\node{i}[u]
	+ \sqrt{\snr_{ii}} \ \channelfd_{ii}[u] \prerffd\node{i} \prebbfd\node{i}[u] \symbvecfd\node{j}[u] + \noisevecfd\node{i}[u] \right) \label{eq:est-i}
	\end{align}
	\begin{align}
	\rxvecfd\node{j}[u] =& \combbfd\ctransnode{j}[u] \comrffd\ctransnode{j} \left(\sqrt{\snr_{ij}} \  \channelfd_{ij}[u] \prerffd\node{i} \prebbfd\node{i}[u] \symbvecfd\node{j}[u]
	+ \noisevecfd\node{j}[u] \right) \label{eq:est-j}
	\end{align}
	\setcounter{equation}{\value{mytempeqncnt}}
	\hrulefill
	\vspace*{4pt}
\end{figure*}

\addtocounter{equation}{1}

Recall that we are considering the network where $i$ is transmitting to $j$ while $i$ is receiving from $k$ (in a \fd fashion). A \mimo formulation per subcarrier gives us an estimated received symbol as \eqref{eq:est-i}.
Notice that the transmitted signal from $i$ intended for $j$ traverses through the \si channel $\channelfd_{ii}[u]$ and is received by the combiner at $i$ being used for reception from $k$. 
The estimated receive symbol at $j$ from $i$ is \eqref{eq:est-j},
where $j$ does not encounter any \si since it is a \hd device. We ignore adjacent user interference between $j$ and $k$---justified by the high path loss and directionality at \mmwave.

%% file: eq-si-channel-rice.tex
\begin{align} \label{eq:si-channel-rice}
	\channel_{ii}[d] &= \sqrt{\frac{\kappa}{\kappa + 1}} \channel^{\textrm{LOS}}_{ii} + \sqrt{\frac{1}{\kappa + 1}} \channel^{\textrm{NLOS}}_{ii}[d]
\end{align}

%% file: sec-contribution.tex
\section{Frequency-Selective Beamforming Cancellation Design} \label{sec:contribution}

\begin{figure*}[!t]
	\normalsize
	\setcounter{mytempeqncnt}{\value{equation}}
	\setcounter{equation}{22}
	\begin{align}
	\prebbfdtilde\node{i}[u] = \entry{\left(\matfd{H}_{\mathrm{des}}[u] \matfd{H}\ctrans_{\mathrm{des}}[u] + \frac{\snr_{ii}}{\snr_{ij}} \matfd{H}_{\mathrm{int}}[u] \matfd{H}\ctrans_{\mathrm{int}}[u] + \frac{\Nrftx\node{i}}{\snr_{ij}}\mathbf{I} \right)\inv \matfd{H}\ctrans_{\mathrm{des}}[u]}{:,0:\Ns\node{i}-1}  \label{eq:rzf}
	\end{align}
	\setcounter{equation}{\value{mytempeqncnt}}
	\hrulefill
	\vspace*{4pt}
\end{figure*}

\addtocounter{equation}{1}

In this section, we present a frequency-selective \bfc design for the system described in Section~\ref{sec:system-model}. 
Our goal is to design precoding and combining strategies across nodes $i$, $j$, and $k$ that enable \fd operation at $i$ by mitigating the \si it would otherwise incur. In this design, we assume perfect \gls{csi} across nodes. There are two primary factors to keep in mind throughout our design. First, the hybrid beamforming architecture involves a baseband and an \rf beamformer at each transmitter and each receiver. Second, being a frequency-selective one, our design will be on a per subcarrier basis: the baseband beamformer is frequency-selective but the \rf beamformer is frequency-flat and would ideally be designed to satisfy all subcarriers to some degree.

\subsection{Frequency-Selective Hybrid Approximation}
We first present a frequency-selective hybrid approximation algorithm that we will use throughout our design. This algorithm extends \gls{omp} based hybrid approximation \cite{omp-heath} from the conventional frequency-flat setting to a frequency-selective one. In \gls{omp} based hybrid approximation, the constraints of the \rf beamformer (e.g., phase quantization and constant amplitude) are captured in a codebook matrix $\Arffd$ whose columns are beamformers satisfying the constraints. Conventional \gls{omp} hybrid approximation is represented in \eqref{eq:omp}, where a fully-digital, frequency-flat beamforming matrix $\xfd$ is being approximated by the product $\xrffd\xbbfd$ for a given codebook $\Arffd$ and number of \rf chains \Nrf. We leave the precise definition of \gls{omp} hybrid approximation to \cite{omp-heath}. In short, \gls{omp} hybrid approximation builds \xrffd column-by-column by choosing the columns in $\Arffd$ that correlate the best with the columns of $\xfd$, appropriately updating its search based on the chosen \rf beamformers as the algorithm progresses. Note that, when transmitting/receiving $\Ns$ streams on $\Na$ antennas, we have $\xfd \in \complexpow{\Na}{\Ns}$, then $\xrffd \in \complexpow{\Na}{\Nrf}$ and $\xbbfd \in \complexpow{\Nrf}{\Ns}$.
\begin{align}
\left(\xrffd,\xbbfd\right) = \text{omp\_hybrid\_approx}\left(\xfd,\Arffd,\Nrf\right) \label{eq:omp}
\end{align}
To extend this conventional frequency-flat \gls{omp} hybrid approximation to a frequency-selective one, we present the following. Given a set of fully-digital beamforming matrices $\left\{\xfd[u]\right\}_{u=0}^{U-1}$, indexed by subcarrier, we construct the matrix $\bar{\xfd}$ by horizontally stacking the matrices as follows.
\begin{align}
\bar{\xfd} =
\begin{bmatrix}
{\xfd[0]} & {\xfd[1]} & \cdots & {\xfd[U-1]}
\end{bmatrix} \in \complexpow{\Na}{U\Ns} \label{eq:fs-omp-stack}
\end{align}
We then invoke conventional \gls{omp} hybrid approximation using \eqref{eq:omp} on $\bar{\xfd}$ instead of $\xfd$, as shown in \eqref{eq:fs-omp}. The effect of this is that the \gls{omp} hybrid approximation algorithm extracts the beamformers in $\Arffd$ that satisfy beamformers across all subcarriers. 
\begin{align}
\left(\xrffd,\xbbfdtilde\right) = \text{omp\_hybrid}\left(\bar{\xfd},\Arffd,\Nrf\right) \label{eq:fs-omp}
\end{align}
The returned baseband matrix  $\xbbfdtilde \in \complexpow{\Nrf}{U\Ns} $ is then of the fashion
\begin{align}
\xbbfdtilde & = 
\begin{bmatrix}
{\xbbfd[0]} & {\xbbfd[1]} & \cdots & {\xbbfd[U-1]} 
\end{bmatrix} \label{eq:fs-omp-stack-bb} , 
\end{align}
where the baseband beamforming matrices per subcarrier are horizontally stacked. The returned \rf matrix $\xrffd$ is of the standard fashion having dimensions number of antennas by number of \rf chains. We refer to this method as frequency-selective \gls{omp} hybrid approximation (FS-OMP) throughout our design.

\subsection{Precoding and Combining at the Half-Duplex Nodes}
Taking the \gls{svd} of the desired channels on subcarrier $u$, we get
\begin{align}
\channelfd_{ki}[u] &= \svdfd{\channelfd_{ki}}{[u]} \label{eq:svd-ki}\\
\channelfd_{ij}[u] &= \svdfd{\channelfd_{ij}}{[u]} \label{eq:svd-ij}, 
\end{align}
where the singular values in both are decreasing along their diagonals. We set the desired precoder at $k$ and the desired combiner at $j$ to the so-called eigenbeamformers
\begin{align}
\prefd\node{k}[u] &= \entry{\rsingfd{\channelfd_{ki}}{[u]}}{:,0:\Ns\node{k}-1} \\
\comfd\node{j}[u] &= \entry{\lsingfd{\channelfd_{ij}}{[u]}}{:,0:\Ns\node{i}-1},
\end{align}
where $\Ns\node{k},\Ns\node{i}$ streams are being communicated on each subcarrier's strongest so-called eigenchannels.
Being fully-digital beamforming matrices, we now seek to represent them in a hybrid fashion. To do this, we employ FS-OMP by building
$\bar{\prefd}\node{k}$ from $\{\prefd\node{k}[u]\}$ and $\bar{\comfd}\node{j}$ from $\{\comfd\node{j}[u]\}$, as described by \eqref{eq:fs-omp-stack}. Invoking \eqref{eq:fs-omp} on each gives us 
\begin{gather}
\left(\prerffd\node{k}, \bar{\prefd}_{\mathsf{BB}}\node{k}\right) = \text{omp\_hybrid}\left(\bar{\prefd}\node{k},\Arffd\node{k},\Nrftx\node{k}\right) \label{eq:fs-omp-k} \\
\left(\comrffd\node{j}, \bar{\comfd}_{\mathsf{BB}}\node{j}\right) = \text{omp\_hybrid}\left(\bar{\comfd}\node{j},\Arffd\node{j},\Nrfrx\node{j}\right), \label{eq:fs-omp-j}
\end{gather}
where $\Arffd\node{m}$ is the beamforming codebook matrix at $m \in \{j,k\}$.
Unpacking $\bar{\prefd}_{\mathsf{BB}}\node{k}$ and $\bar{\comfd}_{\mathsf{BB}}\node{j}$ gives us the baseband beamforming matrix at $k$ and $j$, respectively, for each subcarrier as described by \eqref{eq:fs-omp-stack-bb}. We fix these hybrid beamformers at the half-duplex nodes $k$ and $j$, concluding their design.

\subsection{Precoding and Combining at the Full-Duplex Node}
Having set the precoder at $k$ and the combiner at $j$, we focus our attention to designing the precoder and combiner at $i$. 
We begin with considering the eigenbeamformers associated with receiving from $k$ and transmitting to $j$. Referring to \eqref{eq:svd-ki} and \eqref{eq:svd-ij},
\begin{align}
\comfd\node{i}[u] &= \entry{\lsingfd{\channelfd_{ki}}{[u]}}{:,0:\Ns\node{k}-1} \\
\prefd\node{i}[u] &= \entry{\rsingfd{\channelfd_{ij}}{[u]}}{:,0:\Ns\node{i}-1} .
\end{align}
We invoke FS-OMP to get the hybrid approximations of each and unpack the baseband beamformers per subcarrier to get
\begin{gather}
\left(\comrffd\node{i}, \left\{\combbfd\node{i}[u]\right\}\right) \label{eq:hybrid-combiner} \\ 
\left(\prerffd\node{i}, \left\{\prebbfd\node{i}[u]\right\}\right) \label{eq:hybrid-precoder}.
\end{gather}
Now, we choose to fix the baseband and \rf combiners to those in \eqref{eq:hybrid-combiner} and the \rf precoder to that in \eqref{eq:hybrid-precoder}. 
The final stage of our design is in tailoring the baseband precoder on each subcarrier to avoid introducing \si.

We desire our choice of precoder $\{\prebbfdtilde\node{i}[u]\}$ to be such that the received \si is mitigated on a per subcarrier basis.
We have chosen to fix $\prerffd\node{i}$ due to its constant amplitude constraint and quantized phase shifters. 
Thus, we have set every baseband and \rf beamformer at each transmitter and receiver except the baseband precoder at $i$. 
This motivates us to define the effective \si channel
\begin{align}
\matfd{H}_{\mathrm{int}}[u]  \triangleq \combbfd\ctransnode{i}[u]\comrffd\ctransnode{i} \channelfd_{ii}[u] \prerffd\node{i} \label{eq:eff-int}
\end{align}
and the effective desired channel from $i$ to $j$ as
\begin{align}
\matfd{H}_{\mathrm{des}}[u]  \triangleq \combbfd\ctransnode{j}[u]\comrffd\ctransnode{j} \channelfd_{ij}[u] \prerffd\node{i} \label{eq:eff-des}
\end{align}
With these definitions, we recognize that a desirable design of the baseband precoder at $i$ resembles that of a \rzf transmitter (or commonly called a \lmmse transmitter) \cite{heath_lozano}. In such a precoder design, the signal-to-leakage-plus-noise ratio is the metric of choice, rather than the signal-to-interference-plus-noise ratio which would apply to the combiner design. A \rzf design resembles an \lmmse one, where trading off matched filtering for interference suppression is done dynamically based on the \glspl{snr} involved. We invoke \eqref{eq:rzf}, where ${\prebbfdtilde}\node{i}[u]$ is our baseband precoder at $i$ on subcarrier $u$ with \bfc applied. 

Upon normalizing all our beamformers according to our power constraint, this concludes our design, having set the \rf beamformers at all nodes and the baseband beamformers on all subcarriers at all nodes.

\subsection{Design Remarks}

We note that an \rzf, like an \lmmse, will be driven to a \zf solution when the interference dominates. Such a scenario is common in \fd applications since the \si channel is often much stronger than the desired channels. This indicates that \zf designs like that in \cite{ipr-bfc-2019-globecom} will likely be similar to the solution arrived at by a \rzf design. For this reason, we expect that appropriate dimensionality should be offered to the baseband precoder to allow a \zf solution requiring $\Nrftx\node{i} \geq (\Ns\node{i} + \Ns\node{k})$ to successfully design a precoder at $i$ that avoids the interference in \eqref{eq:eff-int} completely. This is because there are $\Ns\node{k}$ dimensions being received by the combiner
and another $\Ns\node{i}$ dimensions are needed in the null space to transmit on.

%% file: sec-simulation-results.tex
\section{Simulation and Results} \label{sec:simulation-results}

To validate our design, we simulated three scenarios using the following parameters, where each scenario is a variant of the three node network referenced throughout. The number of clusters and rays per cluster is the same for each tap. Ray delays for a given tap are uniformly distributed between sampling instances $(d\Ts)$, where $\fs = 1 / \Ts = 2$ GSPS. For the desired channels, the number of clusters and rays per cluster are uniformly distributed on $[1,6]$ and $[1,10]$, respectively. Each ray's \gls{aod} and \gls{aoa} are Laplacian distributed with a standard deviation of $0.2$ and a mean according to its cluster mean \gls{aod} and \gls{aoa}, which is uniformly distributed on $[0,\pi]$.
We consider a root raised cosine pulse shape where $\beta = 1$.
The maximum delay of frequency-selective channels $D$ is varied throughout our simulated scenarios, though we consistently use an \gls{ofdm} cyclic prefix of length $N_{\textrm{CP}} = D / 4$. 

We transmit two streams on each link, i.e., $\Ns\node{k} = \Ns\node{i} = 2$. Horizontal half-wavelength \glspl{ula} are used at all nodes, where the number of transmit antennas and the number of receive antennas are both $32$. The transmit and receive arrays at $i$ are vertically stacked with a separation of $10$ wavelengths, a factor only of concern to create the spherical-wave \mimo \si channel in \eqref{eq:si-channel-rice}.
For the \si channel, we use a Rician factor of $\kappa = 10$ dB and a \gls{nlos} channel with the same parameters as the simulated desired channels, except with fewer clusters and rays. The number of clusters is uniformly distributed on $[1,3]$, and the number of rays per cluster is uniformly distributed on $[1,6]$. In all simulated scenarios, we let $\snr_{ii} = 80$ dB due to close proximity of the transmit and receive arrays at $i$.


During \gls{omp} hybrid approximation, we use a \gls{dft} codebook matrix across at all nodes. Note that this choice implicitly suggests that the phase shifters have a resolution of at least $2\pi / \Na$, where $\Na$ is the number of antennas at the beamformer being approximated. 

\begin{table}[!t]
	\renewcommand{\arraystretch}{1.3}
	\caption{Simulated Scenarios and their Parameters}
	\label{tab:sim}
	\centering
	\begin{tabular}{|c|c|c|c|c|c|}
		\hline
		Scenario & $U$ & $\Nrftx\node{i}$ & $\Nrfrx\node{i}$ & $\Nrfrx\node{j}$ & $\Nrftx\node{k}$ \\
		\hline
		Equal users, low selectivity & $8$ & $6$ & $2$ & $2$ & $2$ \\
		\hline
		Equal users, high selectivity & $128$ & $8$ & $4$ & $4$ & $4$ \\
		\hline
		Disparate users, low selectivity & $8$ & $6$ & $2$ & $2$ & $2$ \\
		\hline
	\end{tabular}
\end{table}

\subsection*{Important Benchmarks}
Our primary metric for evaluating our design is the sum of the spectral efficiencies from $i$ to $j$ and from $k$ to $i$.


\subsubsection{Ideal Full-Duplex}
To evaluate our design, we compare its performance to that of an ideal \fd system, where the two links simultaneously operate interference-free. There are two ideal \fd scenarios to consider: (i) two non-interfering links operating with fully-digital eigenbeamformers and (ii) two non-interfering links operating with hybrid-approximated eigenbeamformers. We consider both because there will be some loss attributed to frequency-selective hybrid approximation and to \bfc design---plotting them against our achieved results will indicate our design's relative performance.

\subsubsection{Half-Duplex}
For evaluating the \hd spectral efficiency, we again consider both a fully-digital and a hybrid case, both using eigenbeamformers. In both, it is important to note that we assume equal time-sharing of the medium by the two links. (Of course, equal time-sharing may not satisfy a particular application's sense of fairness.) Given equal time-sharing, the fully-digital \hd sum spectral efficiency is half of the fully-digital ideal \fd one. The hybrid \hd sum spectral efficiency is half of the hybrid ideal \fd one. It is the goal of our \fd design to outperform \hd operation.

\subsection*{Scenario \#1: Similar Users in a Mildly Selective Channel}

In this simulated scenario, we have considered equal \glspl{snr} on the links from $k$ to $i$ and from $i$ to $j$, i.e., $\snr_{ki} = \snr_{ij}$. Such a scenario could correspond to a case where $k$ and $j$ are collocated, perhaps together comprising a single \fd node. (However, we remark that we have not considered reciprocal channels.) Additionally, we have considered a mildly selective channel where $D = 8$ and $U = 8$. As indicated by the first row of \tabref{tab:sim}, we have allotted $6$ \rf chains to our precoder at $i$. This provides the dimensionality for sufficient performance of our design, as described in our design remarks, where we will use the additional \rf chains for mitigating \si.

\begin{figure}[!t]
	\centering
	\includegraphics[height=0.36\textwidth]{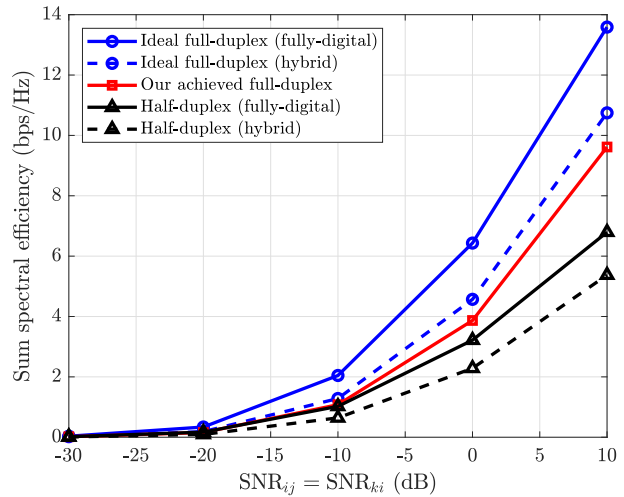}
	\caption{Results of simulating Scenario \#1 showing sum spectral efficiency as a function of $\snr_{ij} = \snr_{ki}$ when $\snr_{ii} = 80$ dB.}
	\label{fig:scenario-1}
\end{figure}

The results of this scenario are shown in \figref{fig:scenario-1}, which shows the sum spectral efficiency of the two links as a function of their \glspl{snr}. 
The two variants of ideal \fd operation are shown, where the fully-digital case outperforms the hybrid one, as expected, due to imperfect hybrid approximation by \gls{omp}. Also included are the fully-digital and hybrid \hd cases. Between \hd and ideal \fd operation lies our achieved \fd sum spectral efficiency.
The goal of \fd operation, of course, is to outperform \hd operation, which is achieved by our design. 
In fact, the gain that can be seen over \hd operation by our design is almost completely attributed to the spectral efficiency of the link from $i$ to $j$ since the link from $k$ to $i$ is relatively unaffected by our design. This is thanks to the \zf nature of the precoder at $i$ given that $\snr_{ii}$ is so strong.

We now comment on our frequency-selective hybrid approximation algorithm based on \gls{omp} (i.e., FS-OMP). In this case we have considered a mildly selective channel, meaning that there are fewer subcarriers. For this reason, it is expected that frequency-selective hybrid approximation is more likely to offer better performance since the frequency-flat \rf beamformer has to satisfy fewer subcarriers. In general, increasing the number of \rf chains will improve the accuracy of hybrid approximation, which is supported in the next simulated scenario. Hybrid approximation (which is inevitably imperfect) will always be outperformed by a fully-digital solution free of constraints. This can be observed in \figref{fig:scenario-1} between the two ideal \fd cases and between the two \hd cases. At best, our design can aim for the performance seen by the hybrid-approximated ideal \fd spectral efficiency. Given all of this, we remark that our frequency-selective \bfc design could be enhanced with improvements to hybrid approximation.



\subsection*{Scenario \#2: Similar Users in a Highly Selective Channel}

In this scenario, we again consider users with equal \glspl{snr}, i.e., $\snr_{ij} = \snr_{ki}$. However, this time, we consider a highly selective channel where the number of channel taps is $D = 128$ and the number of subcarriers is accordingly $U = 128$. Most immediately, we remark that high frequency-selectivity will introduce challenges in hybrid approximation: the frequency-flat \rf beamformers must satisfy many subcarriers. To account for this, we increase the number of \rf chains at each transmitter and each receiver compared to the mildly selective case in Scenario \#1. The parameters used are shown in the second row of \tabref{tab:sim}.

\begin{figure}[!t]
	\centering
	\includegraphics[height=0.36\textwidth]{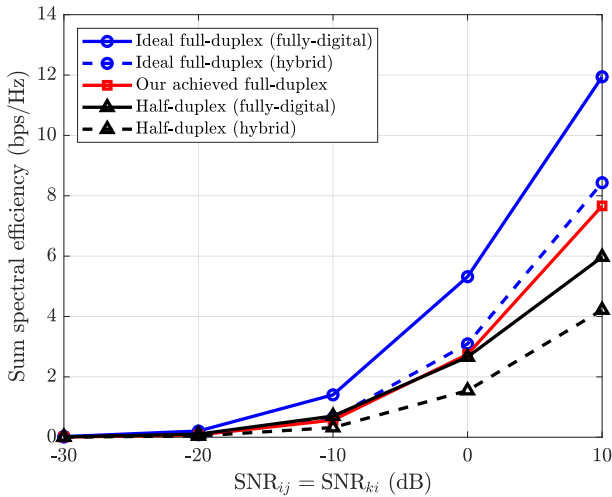}
	\caption{Results of simulating Scenario \#2 showing sum spectral efficiency as a function of $\snr_{ij} = \snr_{ki}$ when $\snr_{ii} = 80$ dB.}
	\label{fig:scenario-2}
\end{figure}

The results of this scenario are in \figref{fig:scenario-2}. Comparing this to the results of Scenario \#1, we make a few points. First, we remark that such high selectivity introduces significant challenges in hybrid approximation, even with the increased number of \rf chains. This introduces sizable gaps between a fully-digital curve and its respective hybrid curve. For this reason, it's important that we compare our design to the \fd and \hd hybrid cases rather than to the fully-digital ones. Our design does quite well in approaching the hybrid ideal \fd spectral efficiency, nearly doubling the hybrid \hd spectral efficiency. These results indicate that our design extends from a mildly selective scenario to a highly selective one quite well.

\subsection*{Scenario \#3: Disparate Users in a Mildly Selective Channel}

In this scenario, we consider the case when the link from $k$ to $i$ is $30$ dB weaker than the link from $i$ to $j$, i.e., $\snr_{ij} = \snr_{ki} + 30 $ dB. Recall that our design sacrifices some performance on the link from $i$ to $j$ while aiming to preserve the link from $k$ to $i$. We consider user disparity to see how this loss on $i$ to $j$ affects the overall performance of our design. The simulation parameters for this can be seen in \tabref{tab:sim}, where we have allotted $6$ \rf chains to the transmitter to exaggerate our point. A lower number of \rf chains at the transmitter yield similar, but less convincing, results.
We remark that we forgo showing the alternative, when $k$ to $i$ is stronger than $i$ to $j$, because such a scenario is favored by our design by preserving the stronger link.

\begin{figure}[!t]
	\centering
	\includegraphics[height=0.36\textwidth]{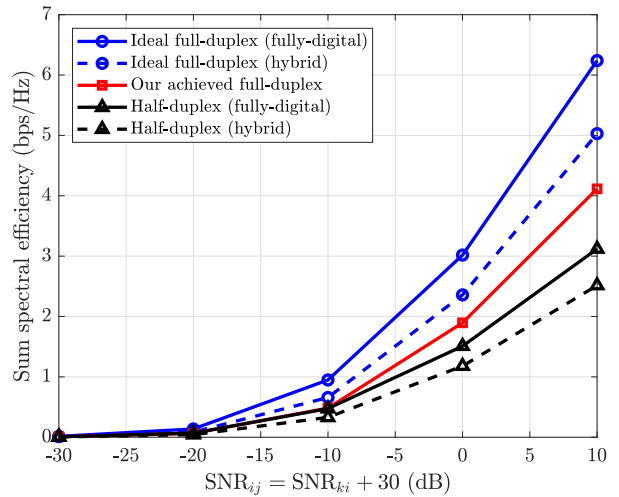}
	\caption{Results of simulating Scenario \#3 showing sum spectral efficiency as a function of $\snr_{ij} = \snr_{ki} + 30$ dB when $\snr_{ii} = 80$ dB.}
	\label{fig:scenario-3}
\end{figure}

The results of this scenario are shown in \figref{fig:scenario-3}, where the sum spectral efficiency is shown as a function of $\snr_{ij}$. 
The achieved sum spectral efficiency during \fd operation with our design is comprised of the degraded link on $i$ to $j$ and the relatively unaffected link from $k$ to $i$. However, since $\snr_{ki}$ is much weaker than $\snr_{ij}$, the losses on $i$ to $j$ are magnified, given our assumption of equal time-sharing for \hd operation. Even with $30$ dB of disparity, the sum spectral efficiency achieved by our \fd design outperforms \hd operation. These results indicate that our design, while somewhat lopsided in that it only tailors precoder in its design, manages to produce meaningful gains over \hd, both fully-digital and hybrid.


%
%
%
%

%% file: sec-conclusion.tex
\section{Conclusion} \label{sec:conclusion}


In this paper, we presented a frequency-selective beamforming strategy that enables simultaneous in-band transmission and reception at a \mmwave transceiver employing fully-connected hybrid beamforming. Our design leverages existing \gls{omp} based hybrid approximation to account for quantized phase shifters and lack of amplitude control associated with \rf beamforming. Our strategy addresses the frequency-selectivity that \mmwave systems are likely see over wideband communication, mitigating \si on each subcarrier while maintaining service to the desired users. Results from simulation indicate that our strategy offers \fd spectral efficiency gains in various channel conditions.

%% file: sec-bibliography.tex
\bibliographystyle{bibtex/IEEEtran}
\bibliography{refs}